\documentstyle[aps]{revtex}
\begin{document}
\wideabs{
\title{The superconducting phase transition and gauge dependence}
\author{Claude de Calan and Flavio S. Nogueira}
\address{Centre de Physique Th\'eorique,  
Ecole Polytechnique, 
F-91128 Palaiseau Cedex, FRANCE}
\date{Received \today}
\tighten
\maketitle

\begin{abstract}
The gauge dependence of the renormalization group 
functions of the Ginzburg-Landau model  
is investigated. The analysis is done by means of the 
Ward-Takahashi identities. 
After defining the local superconducting order parameter, 
it is shown that its exponent $\beta$ is in fact gauge 
independent. This happens because in $d=3$ the Landau gauge is the 
only gauge having a physical meaning, a property not shared by the 
four-dimensional model where any gauge choice is possible. 
The analysis is 
done in both the context of the $\epsilon$-expansion and in the 
fixed dimension approach. It is pointed out the differences that arise 
in both of these approaches concerning the gauge dependence.    
\end{abstract}
\draft
}

Eltitzur's theorem \cite{Elitzur} states that a local gauge symmetry 
cannot be sponteneously broken. As a consequence, no local order 
parameter can be defined in a model described by a locally gauge 
invariant action. However, the gauge symmetry can be explicitly broken, 
for instance, by adding a gauge fixing term to the action. In the 
context of the usual lattice Ginzburg-Landau (GL) model 
\cite{Kleinert} but with a gauge fixing term, Kennedy and King 
\cite{Kennedy} have proposed the following 
{\it non-local} order parameter (OP):

\begin{equation}
\label{G}
G_{\infty}=\lim_{|{\bf x}-{\bf y}|\to\infty}<G({\bf x},{\bf y})>,
\end{equation}
where the operator $G({\bf x},{\bf y})$ is given by a smeared string:

\begin{equation}
G({\bf x},{\bf y})=\phi_{\bf x}\exp\left[-ie\sum_{\mu,{\bf z}}
A_{\mu{\bf z}}h_{\mu{\bf z}}\right]\phi^{*}_{\bf y},
\end{equation}
where $\phi_{\bf x}$ and $A_{\mu{\bf x}}$ are the scalar and gauge 
fields, respectively, defined in a lattice. The gauge group is 
non-compact. The field 
$h_{\mu{\bf z}}=\Delta_{\mu}V_{{\bf z}-{\bf x}}-\Delta_{\mu}
V_{{\bf z}-{\bf y}}$, with $\Delta_{\mu}$ being a lattice derivative and 
$V$ is the kernel of $(-\bar{\Delta})^{-1}$, where 
$\bar{\Delta}$ is the lattice Laplacian. The operator $G$ so defined 
is gauge invariant and in the Landau gauge 
$\sum_{\mu,{\bf z}}A_{\mu{\bf z}}h_{\mu{\bf z}}$ vanishes \cite{Kennedy}. 
By using the so defined order parameter, Kennedy and King have shown 
that there is true long range order for $d\leq 4$ 
($G_{\infty}\neq 0$) only if the 
Landau gauge is fixed. Note that this result was proved for a lattice 
GL model only and is not a trivial matter to extend the analysis of 
Ref. \cite{Kennedy} to the continuum. The aim of this report is to 
provide an analysis of the gauge dependence directly in the continuum. 
Also, we will define a local, gauge invariant OP. To this end, 
we will employ the Ward-Takahashi (WT) identities. The study that will 
be undertaken here was initiated recently by one of us \cite{Nogueira}. 

Let us consider the following bare action for the GL model:

\begin{eqnarray}
\label{action}
S&=&\int d^d x\left[\frac{1}{4}F_{0}^2+(D_{\mu}^0\phi_{0})^{\dag}
(D_{\mu}^0\phi_{0})+\frac{M_{0}^2}{2}A_{\mu}^0 A_{\mu}^0\right.
\nonumber\\
&+&\left.m_{0}^2|\phi_{0}|^2
+\frac{u_{0}}{2}|\phi_{0}|^4\right]+S_{gf},
\end{eqnarray}
where the zeroes denote bare quantities, $F_{0}^{2}$ is a short for 
$F_{0}^{\mu\nu}F_{0}^{\mu\nu}$ and $D_{\mu}^0=\partial_{\mu}+ie_{0}A_{\mu}^0$.
The $S_{gf}$ is the gauge fixing part and is given by

\begin{equation}
S_{gf}=\int d^d x\frac{1}{2a_{0}}(\partial_{\mu}A_{\mu}^0)^2.
\end{equation}
The mass term for the gauge field is introduced in order to avoid 
infrared divergences. The renormalized counterpart of $M_{0}$ tends to 
zero as approaching the critical point more rapidly than $m$ 
\cite{Nogueira} (quantities 
without zeroes are renormalized quantities. Renormalization is 
done in a standard way \cite{ZJ} and the renormalization conditions 
are the same as in Ref. \cite{Nogueira}). 
We note that only averages of 
gauge invariant operators have a physical meaning. Thus, although 
$<\phi>$ could be different from zero due to the presence of terms 
in the action which break explictly gauge invariance, this average is 
surely gauge dependent. Only averages of gauge invariant quantities 
are gauge independent \cite{ZJ,Collins}. For instance, the superfluid 
density $\rho_{s}=<|\phi|^2>$ is gauge independent. We can write 

\begin{equation}
\rho_{s}=<|\phi|^2>=Z_{\phi}^{-1}<|\phi_{0}|^2>.
\end{equation}
where $Z_{\phi}$ is the wave function renormalization of the scalar field. 
Near the critical point we have 
$Z_{\phi}\sim |m|^{\eta}\sim|t|^{\nu\eta}$, where $t$ is the reduced 
temperature and $\eta$ is the fixed point value of the renormalization 
group (RG) function $m\partial\ln Z_{\phi}/\partial m$ (we assume 
the existence of an infrared stable fixed point; for a discussion 
on this point see Refs.\cite{Herbut,Kiometzis,Folk,Tesanovic,deCN} and 
references therein). 
Note that the critical exponents should be the same 
above and below $T_{c}$. This is a consequence of the fact that the 
same counterterms of the symmetric phase can be used when renormalizing 
the broken symmetry theory if a minimal subtraction is used. This fact is 
a consequence of the WT identities \cite{ZJ,Collins}. $Z_{\phi}$ is 
a gauge dependent quantity (by gauge dependence here we mean dependence on 
$a$ and not on $a_{0}$). This means that $<|\phi_{0}|^2>$ should depend on 
$a$ in such a way as to cancel the gauge dependence of 
$Z_{\phi}^{-1}$. Note that $<|\phi_{0}|^2>$ {\it does not} depend on 
$a_{0}$ since it is a gauge invariant quantity. It is easy to show that 
$<|\phi_{0}|^2>\sim|t|^{\nu(d-2+\eta)}$ near the critical point. 
This scaling suggests the 
following definition of a local OP:

\begin{equation}
\label{ordpar}
\Phi=\sqrt{<|\phi_{0}|^2>},
\end{equation}  
having a critical exponent $\beta=\nu(d-2+\eta)/2$. In order to prove that 
$\beta$ is gauge independent we need to prove that $\eta$ is gauge 
independent. 
Note that the other renormalized parameters of the 
model are trivially gauge independent. This statement implies the 
gauge independence of all beta functions (which does not mean 
necessarily that the bare parameters are gauge independent). 

In order to obtain the gauge dependence of $Z_{\phi}$ we will employ 
the following WT identity:

\begin{eqnarray}
\label{eqaux1}
\left(M^2-\frac{1}{a}\partial^2_{z}\right)\partial_{\mu}^{z}
W_{\mu}^{(2)}(z;y,x)&=&ie\left[\delta(y-z)W^{(2)}(z,x)\right.
\nonumber\\
&-&\left.\delta(z-x)W^{(2)}(y,z)\right].
\end{eqnarray}
By using twice Eq. (\ref{eqaux1}) we obtain

\begin{eqnarray}
\label{WT3}
W^{(2)}_{(\partial_{\mu}A_{\mu})^2}(p)&=&2e^2\int\frac{d^d k}{(2\pi)^3}
\frac{a^2}{(k^2+a M^2)^2}\left[W^{(2)}(p+k)\right.\nonumber\\
&-&\left.W^{(2)}(p)\right], 
\end{eqnarray}
where $W^{(2)}_{(\partial_{\mu}A_{\mu})^2}(p)$ is the Fourier transform 
of 

\begin{eqnarray}
W^{(2)}_{(\partial_{\mu}A_{\mu})^2}(x,y)&=&\int d^d z\left[<
(\partial_{\mu}A_{\mu})^2(z)\phi(x)\phi^{\dag}(y)>\right.\nonumber\\
&-&\left.<(\partial_{\mu}A_{\mu})^2(z)><\phi(x)\phi^{\dag}(y)>\right].
\end{eqnarray}
Let us denote the bare counterpart of $W^{(2)}_{(\partial_{\mu}A_{\mu})^2}$ 
by $W^{(2)}_{(\partial_{\mu}A_{\mu}^0)^2,0}$. We have that 

\begin{equation}
\label{eqaux2}
2a_{0}^2\frac{\partial W_{0}^{(2)}}{\partial a_{0}}(x,y)
=W^{(2)}_{(\partial_{\mu}A_{\mu}^0)^2,0}(x,y), 
\end{equation}
where $W^{(2)}_{0}(x,y)=<\phi_{0}(x)\phi_{0}^{\dag}(y)>$ is the bare 
2-point connected correlation function. Eq. (\ref{WT3}) is valid also 
if we replace the renormalized correlation functions by the bare ones and 
the renormalized couplings by their bare counterparts. Using then a bare 
version of (\ref{WT3}) and Eq. (\ref{eqaux2}), we obtain 

\begin{equation}
\label{eqcentral}
\frac{\partial W^{(2)}_{0}}{\partial a_{0}}(p)=
e^2_{0}\int\frac{d^3k}{(2\pi)^3}\frac{W^{(2)}_{0}(p+k)-W^{(2)}_{0}(p)}{
(k^2+a_{0}M^2_{0})^2}.
\end{equation}
Eq. (\ref{eqcentral}) can be rewritten as

\begin{eqnarray}
\label{eqcentral1}
\frac{\partial\ln Z_{\phi}}{\partial a_{0}}W^{(2)}(p)+
\frac{\partial W^{(2)}}{\partial a_{0}}(p)&=&\nonumber\\ 
e^2_{0}\int\frac{d^3k}{(2\pi)^3}\frac{W^{(2)}(p+k)-W^{(2)}(p)}{
(k^2+a_{0}M^2_{0})^2},
\end{eqnarray}
out of which we obtain 

\begin{eqnarray}
\label{gdepZ}
\frac{\partial\ln Z_{\phi}}{\partial a_{0}}&=&
-e^2_{0}\int\frac{d^d k}{(2\pi)^d}\frac{1}{(k^2+a_{0}M^2_{0})^2}\nonumber\\
&=&e^2_{0}a^{\frac{d-4}{2}}_{0}
M^{d-4}_{0}\left(\frac{d}{2}-1\right)C_{d},
\end{eqnarray}
where $\pi/C_{d}=(4\pi)^{d/2}\Gamma(d/2)\sin(\pi d/2)$. Note that we 
have a pole for $d=4$ in the second line of Eq. (\ref{gdepZ}). This is 
a consequence of the logarithmic divergence for $d=4$. In the 
$\epsilon$-expansion the singular part of the different correlation 
functions is isolated as poles in $1/\epsilon$ with $\epsilon=4-d$ and 
the renormalization constants are written as power series in 
$1/\epsilon$. This way of doing the things leads to the determination 
of the critical exponents as power series in $\epsilon$ \cite{ZJ}. 
The physical case of interest in critical phenomena of superfluid and 
magnetic systems corresponds to $\epsilon=1$. 

By integrating the first line of (\ref{gdepZ}) we obtain

\begin{equation}
\label{gdepZ1}
\ln Z_{\phi}(a_{0})=\ln Z_{\phi}(a_{0}=0)-e^2_{0}a_{0}
\int\frac{d^d k}{(2\pi)^d}\frac{1}{k^2(k^2+a_{0}M^2_{0})^2}.
\end{equation}
Since $e^2_{0}a_{0}=e^2a$ and $a_{0}M^2_{0}=a M^2$, 
we can rewrite Eq. (\ref{gdepZ1}) as 

\begin{equation}
\label{gdepZ2}
\ln Z_{\phi}(a)=\ln Z_{\phi}(a=0)-e^2a
\int\frac{d^d k}{(2\pi)^d}\frac{1}{k^2(k^2+a M^2)^2}.
\end{equation}
Let us assume that $Z_{\phi}(a=0)$ has been evaluated as a power 
series in $1/\epsilon$. After regularizing dimensionally the integral 
in (\ref{gdepZ2}), we obtain

\begin{equation}
\label{gdepeta}
\eta_{\phi}(a)=\eta_{\phi}(a=0)-\frac{a f}{2\pi}, 
\end{equation}
which gives the gauge dependence of $\eta_{\phi}$ in the framework of 
the $\epsilon$-expansion ($f=e^2 m^{-\epsilon}$). 
Let us assume that an infrared stable 
fixed point has been obtained, for instance, by ressummation 
methods \cite{Folk}. As $m\to 0$, $f\to f_{*}\neq 0$ (if $\epsilon=1$), 
but $a$ scales as $m^{-1}$ near the fixed point and 
any non-zero $a$ runs away as $m\to 0$. Thus, the only safe way 
towards the charged fixed point is over the line $a=0$, that is, 
the Landau gauge. Note that for the case of interest in particle 
physics, $d=4$, we obtain the same equation as (\ref{gdepeta}). However, 
for $d=4$ any gauge choice is possible since $f_{*}=0$ in this 
case.

In the fixed dimension approach things work differently. For $d=3$ the 
integral in Eq. (\ref{gdepZ2}) is convergent and we can interchange 
the differentiation with respect to m with the integral sign. 
Since $a e^2$ and $a M^2$ are both RG invariants, we obtain 

\begin{equation}
\label{gdepeta1}
\eta_{\phi}(a)=\eta_{\phi}(a=0),
\end{equation}
and we obtain again that the physical gauge corresponds to $a=0$. At 
this point some remarks are in order. First, from Eq. (\ref{gdepeta1}) 
we obtain $\partial\eta_{\phi}/\partial a=0$ while the same is not 
true for the $\eta_{\phi}(a)$ given in Eq. (\ref{gdepeta}). Second, 
Eq. (\ref{gdepeta1}) can be easily checked at 1-loop order. Concerning the 1-loop example, it is 
instructive to ask ourselves what happens in other fixed dimension 
approaches. For instance, we could perform a critical point ($m=0$) 
calculation where the renormalization conditions are defined at  
non-zero external momenta, taking the symmetrical point for functions 
which depend on more than one momentum variable 
\cite{Herbut,deCalan}. In this case the photon mass $M$ is 
unecessary since the non-zero external momenta take care of infrared 
divergences. The 1-loop expression for $Z_{\phi}$ in an arbitrary 
gauge is in this case rather simple and has been calculated by 
Schakel \cite{Schakel}. It turns out in this case 
that $Z_{\phi}$ is independent of $a$ if $d=3$. 

The gauge dependence of $Z_{m}$ can be obtained in an analogous way. 
From  
$m^2=Z_{\phi}Z_{m}^{-1}m_{0}^2$, we obtain that 
$W^{(2)}_{0}(0)=Z_{m}/m_{0}^2$. Using again the bare version of 
(\ref{WT3}), we obtain exactly the same equation as Eq. (\ref{gdepZ}) but 
with $Z_{\phi}$ replaced by $Z_{m}$. This means that the gauge 
dependence of $Z_{m}$ is the same as for $Z_{\phi}$. If we use 
the $\epsilon$-expansion we have that $\eta_{m}-\eta_{\phi}$ is 
gauge independent since the gauge dependence of $\eta_{m}$ will cancel 
exactly the gauge dependence of $\eta_{\phi}$. In fixed dimension 
$d=3$, on the other hand, $\eta_{m}(a)=\eta_{m}(a=0)$. It 
follows that the critical exponent $\nu$ is gauge independent. Since 
$\eta$ is gauge independent, it follows that $\beta$ is gauge 
independent and the order parameter $\Phi$ has a true physical meaning.


\end{document}